# Non-random coil behavior as a consequence of extensive PPII structure in the denatured state

*Aitziber L. Cortajarena¶, Gregg Lois‡, Eilon Sherman§, Corey S. O'Hern‡, Lynne Regan¶,†,*, and Gilad Haran§,**

¶Department of Molecular Biophysics & Biochemistry, †Department of Chemistry, and ‡Department of Mechanical Engineering and Department of Physics, Yale University, New Haven, CT 06520, Phone: 203 432 9843, Fax: 203 432 5175

§Department of Chemical Physics, Weizmann Institute of Science, P.O.B. 26, Rehovot 76100, Israel, Phone: 972 8 9342625, Fax: 972-8-934-2749

E-mail: lynne.regan@yale.edu; Gilad.Haran@weizmann.ac.il

Running title: Compact conformation in a protein's denatured state.

**Abstract**

Unfolded proteins may contain native or non-native residual structure, which has important implications for the thermodynamics and kinetics of folding as well as for misfolding and aggregation diseases. However, it has been universally accepted that residual structure should not affect the global size scaling of the denatured chain, which obeys the statistics of random coil polymers. Here we use a single-molecule optical technique, fluorescence correlation spectroscopy, to probe the denatured state of set of repeat proteins containing an increasing number of identical domains, from two to twenty. The availability of this set allows us to obtain the scaling law for the unfolded state of these proteins, which turns out to be unusually compact, strongly deviating from random-coil statistics. The origin of this unexpected behavior is traced to the presence of extensive non-native polyproline II helical structure, which we localize to specific segments of the polypeptide chain. We show that the experimentally observed effects of PPII on the size scaling of the denatured state can be well-described by simple polymer models. Our findings suggest an hitherto unforeseen potential of non-native structure to induce significant compaction of denatured proteins, affecting significantly folding pathways and kinetics.

# Introduction

During the folding process, proteins reach a defined, unique, native structure from a poorly defined ensemble of unfolded conformations. Although we have an atomistic description of the structures of the native states of a multitude of proteins, much less is known about the unfolded ensemble. The unfolded state is often treated as some version of a random coil, a realistic model of which is the self-avoiding random walk (SARW), in which different chain segments cannot occupy the same volume element. For a SARW, the relationship between the radius of gyration ($R_g$) (or the hydrodynamic radius ($R_h$)), and the length of the chain is given by the expression $R_g \alpha N^\nu$, where $\nu = 0.59$ and N is the number of units in the chain.[1; 2] For many different denatured proteins the experimentally-measured relationship between $R_g$ or $R_h$ and N is consistent with the behavior of a SARW.[3; 4; 5; 6] This observation has been interpreted as support for an unstructured unfolded state. Conversely, there have been several reports that describe the existence of 'residual structure' in the unfolded state [7; 8; 9; 10] which, in general, does not change the global size scaling of the chain.[11; 12] A detailed understanding of the nature of the unfolded state is essential for a complete understanding of protein folding, because any residual structure in the 'unfolded state' will clearly play a role in modulating both the thermodynamics and kinetics of folding, and might affect the formation of alternative structures, such as amyloid.

Repeat proteins are composed of tandem arrays of a small structural motif.[13] they are widespread in nature and in many cases function as mediators of protein-protein interactions. Their simplified modular structures make them ideal systems for understanding basic principles that drive protein folding.[14; 15; 16] The tetratricopeptide repeat is a 34 amino-acid motif in which two anti-parallel α-helices stack together.[17]

We have previously reported the design and characterization of a consensus tetratricopeptide (CTPR) sequence, and have generated a series of proteins with different numbers of tandem repeats of this sequence.[17; 18; 19] Tandem arrays of CTPR units form a regular, superhelical, extended structure (Fig. 1A).[18; 19] It has been suggested that the folding mechanism of these proteins may be different than that of globular proteins.[18] Here we combine spectroscopy and modeling to shed light on the structure of the denatured state of the CTPR proteins. Surprisingly, we find that extensive non-native structure leads to a compact denatured state, with size scaling strongly deviating from that expected for a random coil.

## Results and Discussion

### The denatured state of CTPR proteins is not a random coil

Fluorescence correlation spectroscopy (FCS) probes the hydrodynamic properties of macromolecules by recording fluctuations in the fluorescence they emit [20]. We employed FCS to measure the hydrodynamic radii of CTPR proteins in dilute solutions, in both their folded and unfolded states. For this purpose we labeled CTPR proteins of 2 to 20 repeats (CTPR2 to CTPR20) with the fluorescent probe Alexa Fluor 488 at a unique C-terminal cysteine residue. We first measured the values of $R_h$ for all the different length CTPR proteins in the native state. The measured values of $R_h$ (Table 1) are in very good agreement with values calculated either from crystal structures of CTPR proteins using the program Hydropro,[21] or by treating the folded structures as rods and using an expression for $R_h$ derived for rigid cylinders[22] (Fig. 1A).

The hydrodynamic properties of CTPR2 to CTPR20 were then studied under strongly denaturing conditions at 6M guanidinium hydrocholorie (GuHCl, see Table 1 and Fig. 1B & 1C). We calibrated the FCS instrument and corrected the results for viscosity and refractive index effects using rhodamine 6G as a standard *(See materials and methods)*. The availability of the full series of CTPR proteins allowed us to obtain a scaling relation for $R_h$ within the series. Unexpectedly, we observed that for a given length of protein, the measured $R_h$ is *different* from that anticipated if the polypeptide behaves as a SARW. The denatured state of many proteins of different lengths has been studied using either hydrodynamic methods,[4; 5] or small angle x-ray scattering,[6] to measure $R_h$ or $R_g$ respectively. These data can be fitted to the empirical scaling factors of $R_h \propto N^{0.57}$ and $R_g \propto N^{0.59}$, respectively,[4; 6] where N is the number of amino acids. Clearly these data are in good agreement with the scaling law for a SARW.[1; 2] For the CTPR proteins, however, the relationship between the measured hydrodynamic radii of the unfolded proteins and N, is best described by $R_h \propto N^{0.37}$ ($R_h = (5.49 \pm 0.25) \times N^{0.37 \pm 0.03}$) (Fig. 1C) [*]. Similar results were obtained with urea as denaturant (not shown). Thus, under strongly denaturing conditions the unfolded state ensembles of the CTPR proteins do not behave as a SARW, but are rather more compact, suggesting the presence of significant structure in the unfolded state [†].

---

[*] The series of $R_h$ values of the native proteins could also be fitted to a power law, giving the relation $R_h = 3.12 \times N^{0.41}$. Note that the pre-exponent is much smaller than obtained for the denatured protein series. The rather high exponent is due to the elongated structure of the native CTPRs.

[†] Interestingly, the m-values for folding of the CPTRs, especially the larger ones, are smaller than expected based on calculations of accessible surface areas of random coiled denatured proteins of corresponding sizes (data not shown), also indicating compact structures.

**Abundance of polyproline II structure in denatured CTPRs**

The properties of proteins are determined by their amino acid composition and sequence. We therefore analyzed the sequence of the CTPR proteins and found that the abundance of several amino acids in the CTPR proteins deviates significantly from that of a reference state, the codon usage in all yeast proteins. Notably, amino acids with a high propensity to form polyproline II (PPII) helical structures, such as P, A, G, and Q[23] appear in the CTPR sequence on average 1.8 times more frequently than in the yeast proteome.

Several recent studies have proposed the existence of a significant population of PPII helical conformations in unfolded proteins, leading to the suggestion that the occurrence of PPII structure may be a general feature of the 'unfolded' state.[24; 25; 26; 27] We used circular dichroism (CD) as a sensitive method by which to detect the presence of PPII helix: it is well established that this helix has a distinctive CD spectrum, with a maximum at 229 nm.[24; 28] Fig. 2A shows CD spectra of CTPR8 at different concentrations of GuHCl, from 3.5 M to 6 M. Importantly, at 3.5 M GuHCl the CD signal characteristic of α-helical structure has completely disappeared (data not shown, see also reference [18]). The spectra clearly show a positive peak around 229 nm. The peak intensity increases with increasing GuHCl concentration, reaching a maximum at a concentration of about 4.5-5 M. This concentration is similar to that at which the unfolding titration curves monitored by either CD signal at 222 nm or by $R_h$ also reach a maximum value. Similar spectra were obtained with urea as denaturant (data not shown). CD spectra of PPII helices are also characterized by a negative peak at 210 nm [27], which is very difficult to observe in the presence of GuHCl. We were able to observe the negative peak by taking a CD spectrum with a short path-length cell and a high protein concentration (not shown).

The modular, repeated, character of the CTPR series of proteins greatly facilitates our investigations of the nature of PPII structure in the unfolded state. An important question is whether the PPII structure extends through the entire sequence, or whether is it present in discrete segments. We estimated the fraction of PPII in CTPR proteins by comparing the observed mean residue ellipticity at 229nm ($[\Theta]_{229}$) with $[\Theta]_{229}$ values reported for PPII-containing peptides.[27] All the CTPR proteins, from CTPR2 to CTPR20, show a substantial PPII content, which we roughly estimate to be as much as ~50 % at 6 M Gu-HCl. The observed $[\Theta]_{229}$ for PPII is the same, regardless of the number of repeats in each protein, indicating that there is an equal amount of PPII structure formed per each repeat, *and* that there is no cooperativity between repeats in formation of PPII structure.

CTPR1, however, does *not* exhibit significant PPII structure under denaturing conditions. CTPR1 has the sequence A-B-$A_{cap}$, where A and B are the two $\alpha$-helices forming the basic repeat unit, and $A_{cap}$ is a solvating helix (for sequence information see *Materials and Methods*),[17] whereas all the longer CTPR proteins have the sequence (A-B)$_n$-$A_{cap}$.[17; 19] CTPR1 thus lacks the B-A inter-repeat sequence. Based on the method of reference [29], this sequence (but not the B-$A_{cap}$ sequence) can in fact be predicted to have high propensity for PPII formation. To test this hypothesis, we synthesized a peptide corresponding to the B-A sequence. This peptide does indeed adopt PPII structure with $[\Theta]_{229}$ value comparable to that of the CTPR proteins. Moreover, PPII formation by the peptide shows the same denaturant concentration dependence as that observed for the full-length proteins (Fig. 2B). This result confirms that the B-A sequence in the CTPR proteins is responsible for the PPII structure and that the PPII segments form independently. Because the B-A peptide is unstructured in the absence of denaturant (Not shown), our results clearly indicate that GuHCl not only promotes formation of PPII structure by destabilizing the native state, but also directly stabilizes this structure.

Finally, the PPII content of peptides has been shown to decrease when the temperature increases.[30] We therefore tested the effect of temperature on the PPII content of highly-denatured CTPR proteins. We observed a reversible loss of PPII structure with increasing temperature, such that the signal at 98ºC indicates essentially zero PPII content (Inset to Fig. 2A).

**Lattice simulations of denatured CTPRs**

To better understand the effect of PPII structure on the dimensions of the denatured CTPRs, we performed simulations using simple polymer models. First, we studied the behavior of a freely-jointed, self avoiding, random walking polymer on a 3-dimensional lattice for polymers with N equal-sized links of length L, where N varies between 10 and 1000, and L is the lattice constant *(see Materials and Methods)*. To allow comparison with the experimental system, we considered any polymer of length N as containing N/34 repeats. For each of these chains, we calculated the $R_g$ and, as expected for a random walk with excluded volume, a log-log plot of $R_g$ *vs.* N yielded a straight line with slope 0.59 (Fig. 3A). To mimic the effect of different lengths of PPII structure within a repeat, we then included in the model links with two different sizes, L (the original) and $N_pL$ (to represent the number of links in the PPI segments), and allowed $N_L$ steps of size $L$ and one step with size $N_pL$ for each repeat. We set the repeat length to be 34 ($N_L + N_p = 34$), as above. This is a reasonable way to model the PPII segments. It is accepted that at least up to a length of 12 residues, PPII peptides may behave as rigid rods,[31; 32] and while isolated non-proline PPII peptides have been shown to be more flexible,[33] there is no such information on non-proline PPII peptides buried within proteins. Note that this scheme creates an *ordered sequence* of rigid and flexible chain stretches. We calculated $R_g$ as a function of N for all $N_p$ values from 1 to 34 and calculated the scaling exponent ν from the slope of log-log plots of $R_g$ *vs.* N. In Fig. 3B we show ν *vs.* $N_p$; ν = 0.59 for $N_p$ =1 (*i.e.* no PPII and the expected random walk with excluded volume behavior). As $N_p$ increases, ν decreases and reaches a minimum when $N_p/34$ is approximately 1/2. The long segments bring the chain to relatively empty regions of the lattice, so that the subsequent steps with small segments are less affected by the excluded volume constraint. This mechanism causes ν to decrease toward the Gaussian coil result of 0.5. As $N_p$ increases further, ν returns to 0.59 and we again have a SARW, but with a larger link size. Note that the scaling exponent ν is smallest when the fraction of PPII segments is 50%, which agrees with experimental estimates of the fraction of PPII residues in each CTPR repeat unit. Although we cannot expect to make fully quantitative comparisons between the results of our simplified simulations and the experimentally observed data for a real polypeptide chain, these results provide a rationale for why the scaling exponent observed for denatured states containing PPII is not the same as that for a random coil, based simply on the intrinsic behavior of mixed polymers.

Interactions between PPII segments are likely to also contribute to the small scaling exponent ν ~ 0.37 found in our experiments.[34] We therefore introduced effective attractive interactions between PPII segments in the simulations by minimizing the distance between a fraction (f; for $0 \leq f \leq 0.8$) of the segments and all other segments in the chain. Incorporating this additional feature into the model has a dramatic effect in reducing the value of $\nu$ still further (Fig. 3C and 3D). Our results suggest that rod-like PPII segments and the interactions between these segments can significantly influence the structure of the unfolded state.

**Conclusion**

In conclusion, we used FCS to measure the value of $R_h$, under denaturing conditions, for a set of repeat proteins with 2 to 20 tandem copies of the same sequence. Our results revealed that the scaling behavior of $R_h$ with N was *not* as expected for a SARW ($R_h \propto N^{0.37}$ *vs.* the expected $R_h \propto N^{0.59}$), indicating a significantly more compact structure. Further investigations showed that there was considerable PPII structure in the denatured proteins. Moreover, we identified the segment of the repeat that formed a PPII helical structure, and showed that PPII was present as regularly spaced, discrete segments throughout the protein. Using simple lattice models of polymers, we showed that chains with a regularly-ordered sequence of short and long links (the latter representing PPII segments) may attain a more compact structure than that of the standard SARW. The incorporation of interactions between PPII segments in the model leads to further compaction of the chains, with size scaling similar to that observed in the experiment. These results present the most comprehensive picture to date of the effect of non-native PPII structure on the nature of the unfolded state ensemble. The formation of non-native structure in the denatured state may have implications for the folding mechanism of CTPR proteins, which we are now exploring[35]. Finally, since proline-rich segments have been recently shown to be effective in counteracting amyloid fiber formation,[36] we hypothesize that PPII structure may in general serve to modulate the aggregation of unfolded proteins.

## Materials and Methods

### Proteins

CTPR proteins are constructed as a tandem array of different numbers of a 34 amino acid consensus CTPR sequence (sequence of the A-B helices: AEAWYNLGNAYYKQGDYDEAIEYYQKALELDPRS). An additional solvating helix is added to the C-terminus of each protein. This solvating helix ($A_{cap}$, AKQNLGNAKQKQG) is similar to the A helix but with some hydrophobic residues substituted by hydrophilic residues. The B-A peptide sequence corresponds to a 34 amino acid consensus TPR in which the B helix precedes the A helix (Ac-YDEAIEYYQKALELDPRSAEAWYNLGNAYYKQGD-$NH_2$).

### Preparation of labeled proteins

The cysteine-reactive maleimide form of Alexa-488 (Molecular Probes, Inc.) was used to label a unique cysteine inserted at the C-terminus of the proteins CTPR2, CTPR3, CTPR4, CTPR6, CTPR8 and CTPR20. The proteins (100 μM) were reduced by incubation during 15 minutes with 10 mM fresh 1,4-dithio-DL-treitol (DTT). DDT was removed by buffer exchange over a PD-10 column (Amersham Bioscience, Uppsala, Sweden) against 500 mM NaCl, 50 mM phosphate, pH 7.0. A solution of Alexa 488 maleimide in water was added drop-wise into the freshly reduced proteins in presence of 1 mM TCEP and with constant stirring to a final concentration of dye equal to 0.5 mM. The reactions were allowed to proceed for 4 hours at room temperature under continuous stirring, under a non-reducing atmosphere of $N_2$ in the dark. After 4 h an excess of β-mercaptoethanol (5% v/v) was added to consume the remaining of thiol reactive reagent during for 15 minutes at 4°C.

The free dye was removed from the reaction mixtures by two consecutive buffer exchange steps over PD-10 column (Amersham Biosciences) followed by concentration and extensive washing on a Centriprep YM-3 filtration unit (Millipore), using a pH 6.8 buffer with 150 mM NaCl, and 50 mM phosphate (PBS).

The extent of labeling was estimated from the absorption spectra of the proteins using the extinction coefficient for Alexa 488 $\varepsilon_{493}$=72,000 $M^{-1}$ $cm^{-1}$ from the Molecular Probes Handbook. The protein concentration was determined using an extinction coefficient calculated from the amino acid composition.[37] The absorbance at 280 nm was corrected for the contribution of the dye.

The protein samples were analyzed by mass spectrometry to confirm the labeling of the protein. Matrix-assisted laser desorption ionization-time of flight (MALDI-TOF) analysis was performed on a Voyager-DE PRO Biospectrometry workstation (AB Applied biosystems, Foster City, CA). The protein samples were mixed with nine volumes of the matrix solution (sinapinic acid 10 mg/ml, 50% acetonitrile, 0.1% trifluoro-acetic acid) and spotted onto the MALDI plate. The labeled proteins were frozen and stored in the dark at -80°C.

**FCS measurements**

FCS measurements were carried out using a laser beam at 488 nm from an Argon-ion laser (Spectra Physics Lasers, Mountain View, CA) as excitation source. The collimated laser beam was focused into the sample, which was mounted on a home-built microscope, by a 60x, 1.2 NA water immersion objective (Olympus, Tokyo, Japan). The laser power intensity was adjusted to 70 µW before the microscope. The beam was directed into the objective by a dichroic mirror (500 DCLP, Chroma, Rockingham, VT) and focused 20 µm into the sample. The fluorescence signal was collected through the same objective and transmitted by the dichroic mirror onto a long-pass filter to eliminate any reflected excitation light. It was then focused onto a 50 µm pin-hole to eliminate out-of-focus photons and divided onto two single-photon avalanche photodiode units (APDs) (Perkin-Elmer Photoelectronics, Fremont,CA) by a 50/50 non-polarizing beam splitter (Unice E-O Services, Chung Li, Taiwan). The signals of the APDs were recorded in the cross-correlation mode for one hour using a Flex02-12D digital real time correlator (Correlator.com, Bridgewater, NJ).

All measurements were carried out on protein samples diluted in PBS buffer to a final concentration of 10 nM, which were transferred into cells constructed from two cover slides and sealed.

Cross-correlation functions were calculated from the signals of the two detectors. The correlation functions presented a fast exponential decay component in the nanosecond to microsecond time scale. In order to calculate the diffusion time of the proteins the correlation curves were fitted using the data from 20 microseconds on, where there is no contribution of photophysical processes, using the following equation:[20]

$$G(\tau) = \frac{1}{N}\left(1 + \frac{t}{\tau_D}\right)^{-1}\left(1 + \frac{t}{\tau_D(z_0/r_0)^2}\right)^{-\frac{1}{2}} \quad (1)$$

where N is the number of molecules in the focal volume, $\tau_D$ is the diffusion time in the radial direction and $r_0$ and $z_0$ are the radial and axial dimensions of the Gaussian observation volume, respectively. The $r_0/z_0$ ratio was obtained for each experimental condition from a measurement of the dye rhodamine 6G (Rh6G), whose diffusion coefficient is known ($D_{Rh6G}$= 280 µm$^2$ s$^{-1}$).[38],[‡] Thus, FCS curves were fitted with two free parameters for the diffusion term ($\tau_d$ and $N$). The diffusion coefficient for each sample was calculated from the $\tau_d$ using the following equation:

$$\tau_D = \frac{r_0^2}{4D_t} \quad (2)$$

using $r_0$ values obtained from the Rh6G measurement. The Stokes-Einstein (SE) relation was used to calculate the hydrodynamic radius ($R_h$) of the molecules from their diffusion coefficient ($D_t$) at a given viscosity ($\eta$):

---

[‡] A recent measurement of the diffusion coefficient of R6G [39] gave a value which is larger by 30% then the above-mentioned number. However, a survey of literature values of protein hydrodynamic radii (e.g. in ref. [40]) and our own measurements [41] both suggest that using the new value will lead to hydrodynamic radii which are consistently smaller not only from calculations directly based on PDB structures, but also from measurements with other methods, such as dynamic light scattering.

$$D_t = \frac{k_B T}{6\pi\eta R_h} \quad (3)$$

where $k_B$ is the Boltzmann constant and T is the absolute temperature.

We also fitted some of the cross-correlation functions over the full time scale including two independent exponentials decays in the equation to consider the triplet state of the dye and the fast process observed in the nanoseconds to microsecond time scale.

$$G(\tau) = \frac{1}{N} \frac{(1 + A_1 \exp(-t/\tau_1) + A_2 \exp(-t/\tau_2))}{\left(1 + \frac{t}{\tau_D}\right)\sqrt{1 + \frac{t}{\tau_D (z_0/r_0)^2}}} \quad (4)$$

where $\tau_1$ and $\tau_2$ are the relaxation times of the two process and $A_1$ and $A_2$ their amplitudes. In this case the FCS curves were fitted with two free parameters for the diffusion term ($\tau_d$ and $N$) and four parameters for the two exponential decays ($A_1$, $A_2$, $\tau_1$ and $\tau_2$). The diffusion time results obtained from the fit of the complete correlation function were in agreement with the values obtained from the fit of only the diffusion component of the function.

In order to obtain a well-behaved standard for the assessment of the effect of refractive index changes on apparent diffusion coefficients, we studied the diffusion of Rh6G molecules in all GuHCl solutions. In a simple liquid, the diffusion coefficient is inversely proportional to the viscosity of the medium as described by SE relation (Equation 3). The SE relation implies that the diffusion time relative to that in water, $\tau/\tau_w$ should be proportional to the viscosity relative to water viscosity, $\eta/\eta_w$, with a proportionality constant of 1. Fig. 5 shows experimental results from a series of measurements in GuHCl solutions. The full line in this figure is a fit to *y=ax* which gives a slope of 0.98, in close agreement with the theoretical SE prediction with a slope of 1. Clearly the SE relation holds rather well in all GuHCl solutions, showing that refractive index changes or other optical effects do not significantly affect the measured diffusion times. We also controlled for possible effects of laser power on measured diffusion times[42; 43] by repeating some of our measurements at a series of lower powers down to 5 μW. We found slight changes in diffusion time values, which are reflected in error values reported in the paper (see Results). Such changes did not affect to any significance the scaling of Rh with size, which is the main observation in this paper.

**Calculation of $R_h$ from crystal structures**

The theoretical diffusion coefficients and the hydrodynamic radius for the different length CTPR proteins were calculated from the x-ray crystal structure coordinates using the program Hydropro version 7c[21] (http://leonardo.fcu.um.es/macromol/programs/hydropro/hydropro.htm). We used the structures of CTPR8 (PDB_ID: 2AVP)[18] and CTPR20 (Kajander et al. submitted paper) previously solved by our group. From the CTPR8 structure we built atomic models for the smaller CTPRs.

Hydropro calculations were performed using the following values in the input file: 3.1 Å for the radius of atomic elements (AER) and a minibead radius from 1 to 5 Å (Sigmin and Sigmax).

**Calculation of $R_h$ with rod-like model**

Because the crystal structures of long CTPR proteins have a extended rod-like shape, we also calculated their hydrodynamic radii using an expression derived by Ortega and Garcia de la Torre for cylindrical particles.[22]

$$R_h = R_h^0 (1.009 + 1.395 \times 10^{-2} (\ln p) + 7.8880 \times 10^{-2} (\ln p)^2 + 6.040 \times 10^{-3} (\ln p)^3) \quad (5)$$

In this expression $R_h^0 = L(3/16 p^2)^{1/3}$ is the hydrodynamic radius of a sphere having the same volume as the cylinder and $p = L/d$ is the aspect ratio between the cylinder length ($L$) and diameter ($d$). Using the dimensions of the TPR superhelix obtained from the crystal structure of CTPR8 (L=72 Å/8 repeats and d=38 Å), we calculated the hydrodynamic properties of the CTPRs of different length.

**CD spectroscopy**

All CD experiments were performed using an AVIV Model 215 CD spectophotometer (AVIV Instruments, Lakewood, NJ) in 150 mM NaCl, 50 mM phosphate buffer pH 6.8 at 25ºC. CD spectra of CTPR proteins were acquired at 12.5 µM protein concentration and at 30 µM for the B-A peptide in a 0.2 cm path length cuvette. The CD spectra were recorded with a band width of 1 nm at 1nm increments and 10 second average time. Molar ellipticity ([Θ]) was calculated using the following equation:

$$[\Theta] = \frac{\theta}{10cnl} \tag{6}$$

where $\theta$ is the ellipticity measured in millidegrees, $c$ is the molar concentration of the sample (M), $n$ is the number of amino acids in the protein and $l$ is the pathlength in cm.

The estimation of the polyproline II helix (PPII) content from the ellipticity at 228 nm was based on proposed ellipticity values for peptides presenting 100% and 0% PPII helix content.[27]

Thermal denaturation was monitored at a protein concentration of 12.5 μM by following the ellipticity signal of the PPII at 228 nm from 10ºC to 98ºC and in the reverse direction from 98ºC to 10ºC in a 0.2 cm path-length cuvette. The temperature ramp was performed in 1ºC steps with an equilibration time at each temperature of 1 min.

**Simulations of self avoiding random walking chains on a 3-dimensional lattice**

Simulations of a random walking chain with excluded volume (SARW) were conducted on a three dimensional cubic lattice with N monomers. To construct a SARW polymer, each monomer was positioned on a lattice site adjacent to the previous monomer with the constraint that no two monomers can occupy the same lattice site. A polymer chain is complete when N monomers have been placed on the lattice in this self-avoiding fashion. The radius of gyration $R_g$ was calculated as an ensemble average over $10^6$ SARW realizations for each N [44], and this data was used to compute the scaling exponent $\nu$ defined through the relation $R_g \propto N^\nu$.

The effect of internal structure on the scaling exponent $\nu$ was investigated by including the statistical effects of rigid polyproline segments. Motivated by experimental findings, $N_p$ consecutive monomers in the middle of each repeat unit were assigned to represent the polyproline segment, with $1 \leq N_p \leq 34$ (where 34 is the number of monomers in each repeat unit). Monomers that are not in the polyproline segment were placed randomly, as in the SARW. The $N_p$ monomers in the polyproline segment were placed such that they all lie in a straight line in one of the six lattice directions and occupy open lattice positions. The orientation of the segment was chosen randomly from the set of all self-avoiding orientations. This model is an extension of the SARW and each repeat unit starts with (34-$N_p$)/2 small steps, followed by one large step of size $N_p$, and ending with (34-$N_p$)/2 small steps. This pattern was continued in the simulations over the entire experimental range of 2-20 repeat units. For each value of $N_p$, the radius of gyration $R_g$ was averaged over an ensemble of $10^6$ realizations and a power law relation was observed between $R_g$ and the number of repeat units. The scaling exponent $\nu$ was calculated for each Np from this data.

The effect of interactions between polyproline segments on the exponent $\nu$ was modeled by randomly choosing a fraction f of the polyproline segments to have an effective attraction with other polyproline segments. In our model, if a segment is chosen, its orientation is not assigned randomly. Rather, for each possible self-avoiding orientation, the distance $r_i$ between the center of the chosen segment and the center of all other existing polyproline segments is calculated. The orientation of the chosen segment is then assigned to be the one that minimizes the sum $\Sigma_i(r_i^2)$. The radius of gyration $R_g$ was measured and averaged over an ensemble of $10^5$ realizations. A power law relation between $R_g$ and number of repeat units was observed and $\nu$ was calculated for $0 \leq f \leq 0.8$.

**Acknowledgements**

We thank Professors A. Horovitz, A.D. Miranker and A. Minsky, as well as members of the Regan and Haran Labs for comments and suggestions on the manuscript. ALC was a recipient of an EMBO short-term fellowship. This research is made possible in part by the historic generosity of the Harold Perlman Family, as well as by partial financial support of the US-Israel binational science foundation (grant no. 2002371 to GH), the NIH (grant no. 1R01GM080515-01 to GH) and the Human Frontier Science Program (to GH and LR). Financial support for GL and CSO from NSF grant number DMR-0448838 is gratefully acknowledged.

**Figure Legends**

**Figure 1.** Rh values of native and unfolded CTPR proteins measured by FCS. A. Correlation between the Rh values for CTPRs of different lengths (CTPR2, CTPR3, CTPR4, CTPT6, CTPR8 and CTPR20) measured by FCS and the Rh values calculated from the crystal structures of the CTPRs using the program Hydropro,[21] (filled circles) or the Rh values calculated using a model that describes the hydrodynamic properties of cylindrical particles (empty circles)[22] (Equation 5 in Materials and Methods). The figure shows the ribbon representations of the structures of two CTPR proteins (CTPR4 and CTPR20). B. FCS cross-correlation curves for the CTPR3 protein at various Gu-HCl concentrations, from 0 to 6M. Arrow indicates the direction of increase in denaturant concentration, which shifts the curves to the right. C. A log-log plot of Rh values of the different CTPRs under strongly denaturing conditions at 6 M Gu-HCl (filled dots) and the least-squares fits to the data (solid line). The red line represents the empirical relation between the measured Rh and protein length for a set of highly denatured globular proteins [4]. The inset shows the dependence on the Gu-HCl concentration of the Rh value for CTPR proteins of different lengths in the unfolded state, CTPR2 (●), CTPR3 (○), CTPR4 (▼), CTPR6 (∇), CTPR8 (■) and CTPR20 (□), indicating that above ~4M Gu-HCl the size of these proteins doesn't change anymore.

**Figure 2.** Circular dichroism spectra of CTPR8 and B-A peptide showing PPII helical structure. A. CD spectra of the unfolded state of CTPR8 at increasing Gu-HCl concentrations. The inset illustrates the temperature dependence of the CD signal, indicating a non-cooperative but reversible melting of the PPII structure. It shows the CD spectra of CTPR8 at 6 M Gu-HCl at 25°C (solid line), at 98°C (dashed and dotted

line) and at 25°C after the sample was heated and cooled down (dashed line). B. CD spectra of the B-A peptide in a 0-6.5 M range of Gu-HCl concentrations.

**Figure 3.** A polymer model for the effect of PPII on the hydrodynamic properties of the denatured state. A. A $\log_{10}$-$\log_{10}$ plot of $R_g$ vs N for the SARW simulation. The ensemble averaged radius of gyration $R_g$ is plotted against the number of monomers N in a classic excluded volume random walking chain (SARW). The data is fit to a power law relation $R_g \propto N^\nu$, and we calculate $\nu = 0.587 \pm 0.003$. The measured exponent is consistent with the accepted value of 0.59. B. The scaling exponent $\nu$ as a function of rigid segment (polyproline) length $N_p$. Two dimensional representations of the polymer chains are pictured for $N_p$ = 4, 17, and 30. Note that the ratio between large (red) and small segments (black) is drawn to scale for each value of $N_p$, but the overall size of the schematic is made smaller for larger $N_p$ to fit into the figure. The value of $\nu$ exhibits a minimum at $N_p$ = 14, roughly half of a repeat unit. Error bars on the values of $\nu$ are $\pm$ 0.003. C. The scaling exponent $\nu$ as a function of polyproline length $N_p$ and several fractions f of polyproline segments that experience an effective attraction. The data for f=0 (no attraction) is identical to Fig. 3B. As f increases, the value of $\nu$ decreases and the location of the minimum moves to larger $N_p$. Circles represent f=0, squares f=0.2, diamonds f=0.4, triangles f=0.6, and stars f=0.8. Error bars on the values of $\nu$ are $\pm$ 0.003 for f=0 and $\pm$ 0.011 all other f. D. Visualization of the polymer chain with N= 204, Np=18, and f=0.6 for a representative configuration from the SARW ensemble.

**Figure. 4.** Dependence of the relative translational correlation time of Rh6G molecules on relative viscosity in 0 to 6 M GuHCl solutions (filled circles). The line

corresponds to the fit of the data to the equation $y=ax$, which gives a slope of 0.98, in agreement with the Stokes-Einstein prediction.

Table

# Table 1: Rh values of native and unfolded CTPR proteins

| # repeats | # aa | Rh (Å) Native | Rh (Å) Unfolded |
|---|---|---|---|
| 2 | 88 | 20.21±0.87 | 29.31±1.21 |
| 3 | 122 | 22.40±0.75 | 34.07±0.31 |
| 4 | 156 | 24.89±0.37 | 36.48±0.42 |
| 6 | 224 | 28.19±1.11 | 41.49±1.12 |
| 8 | 292 | 30.62±0.79 | 45.30±0.81 |
| 20 | 700 | 48.36±0.98 | 64.89±1.32 |

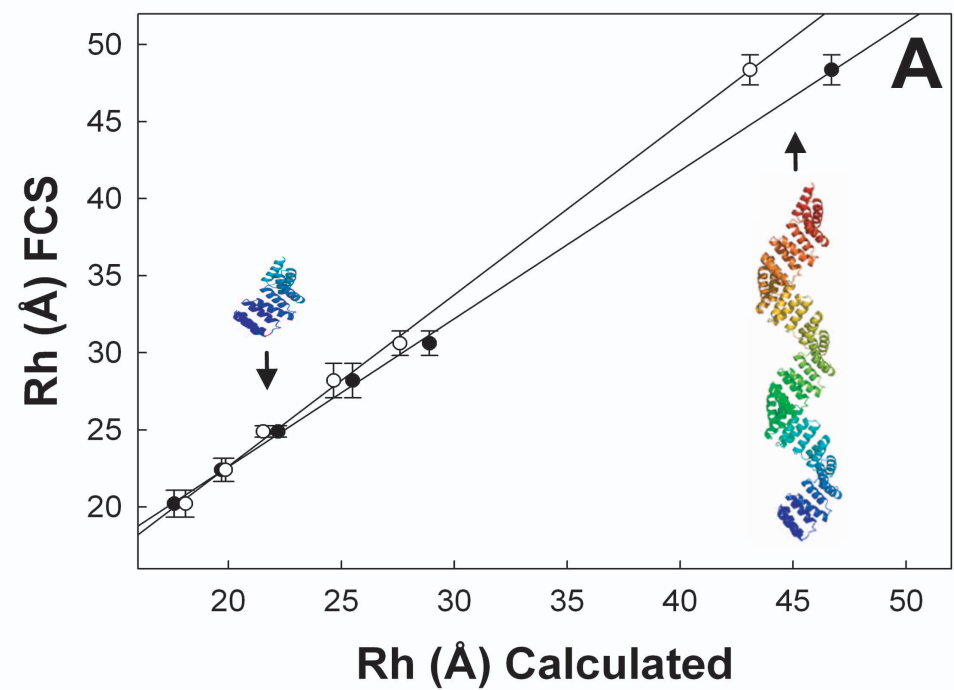

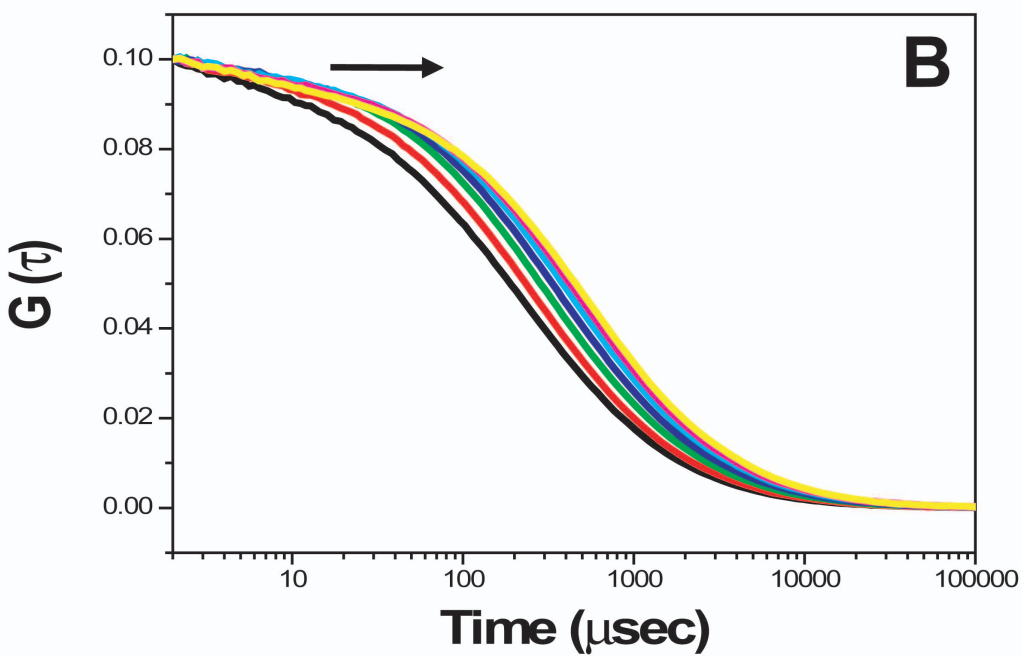

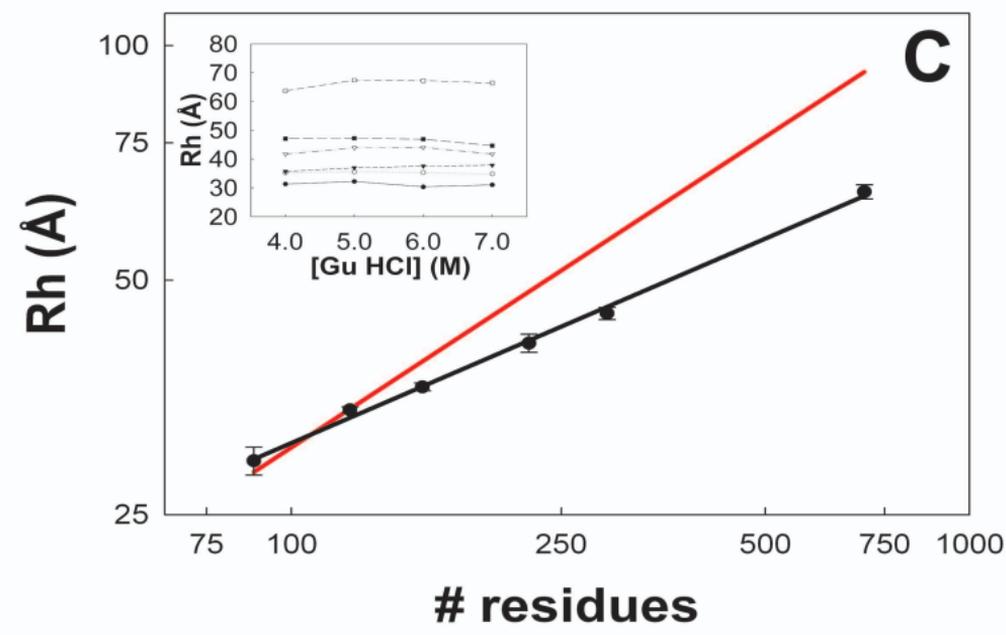

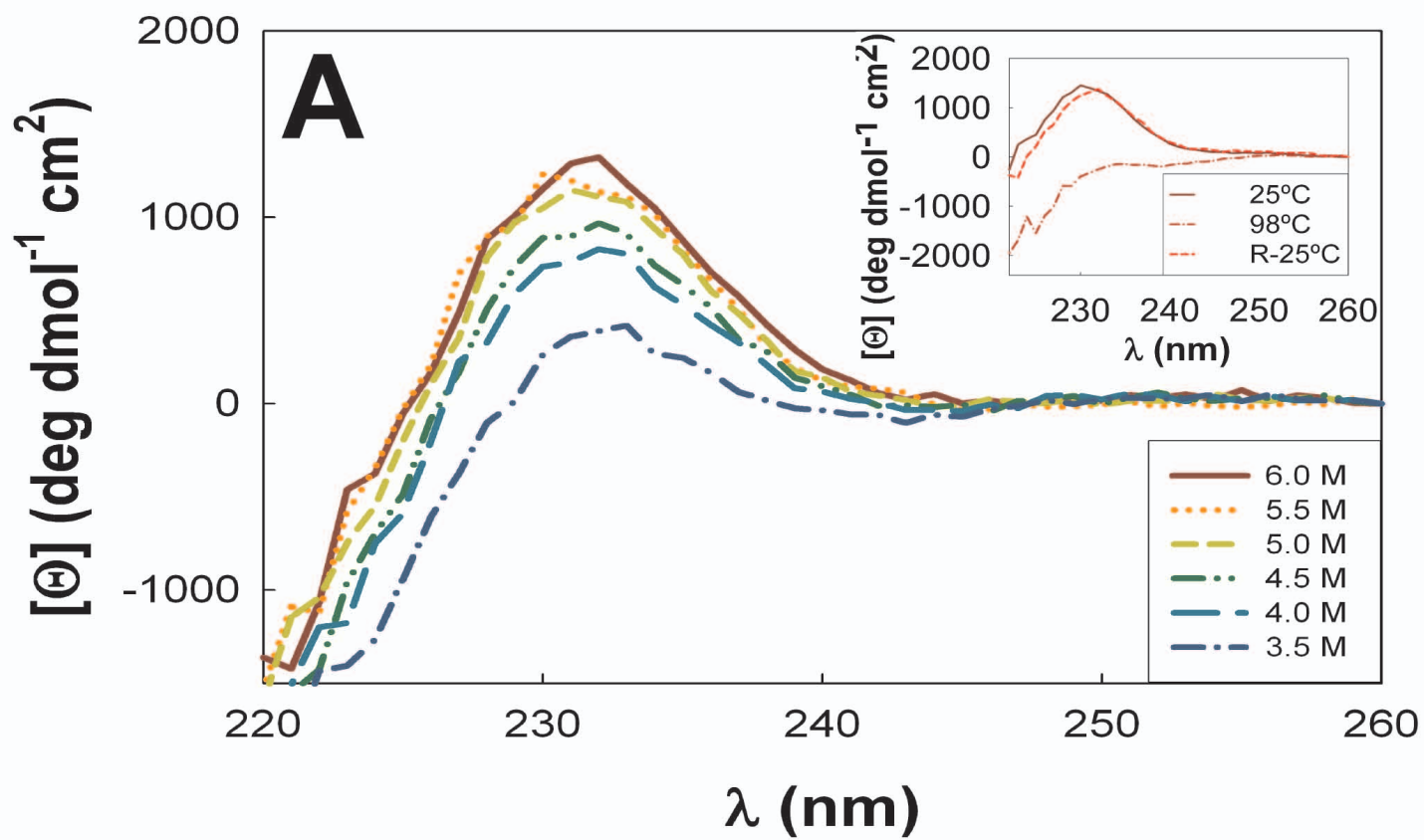
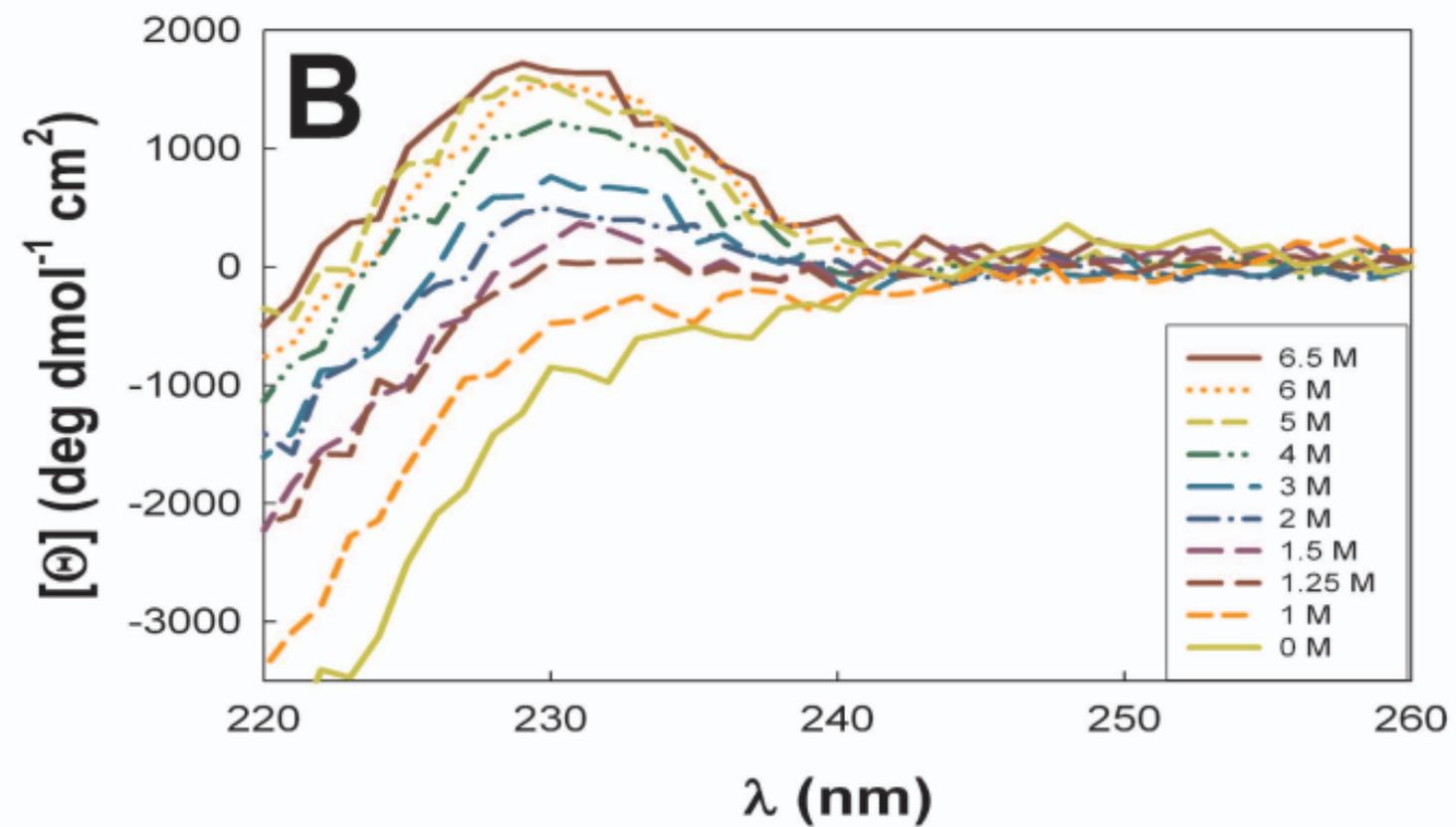

Figure 4

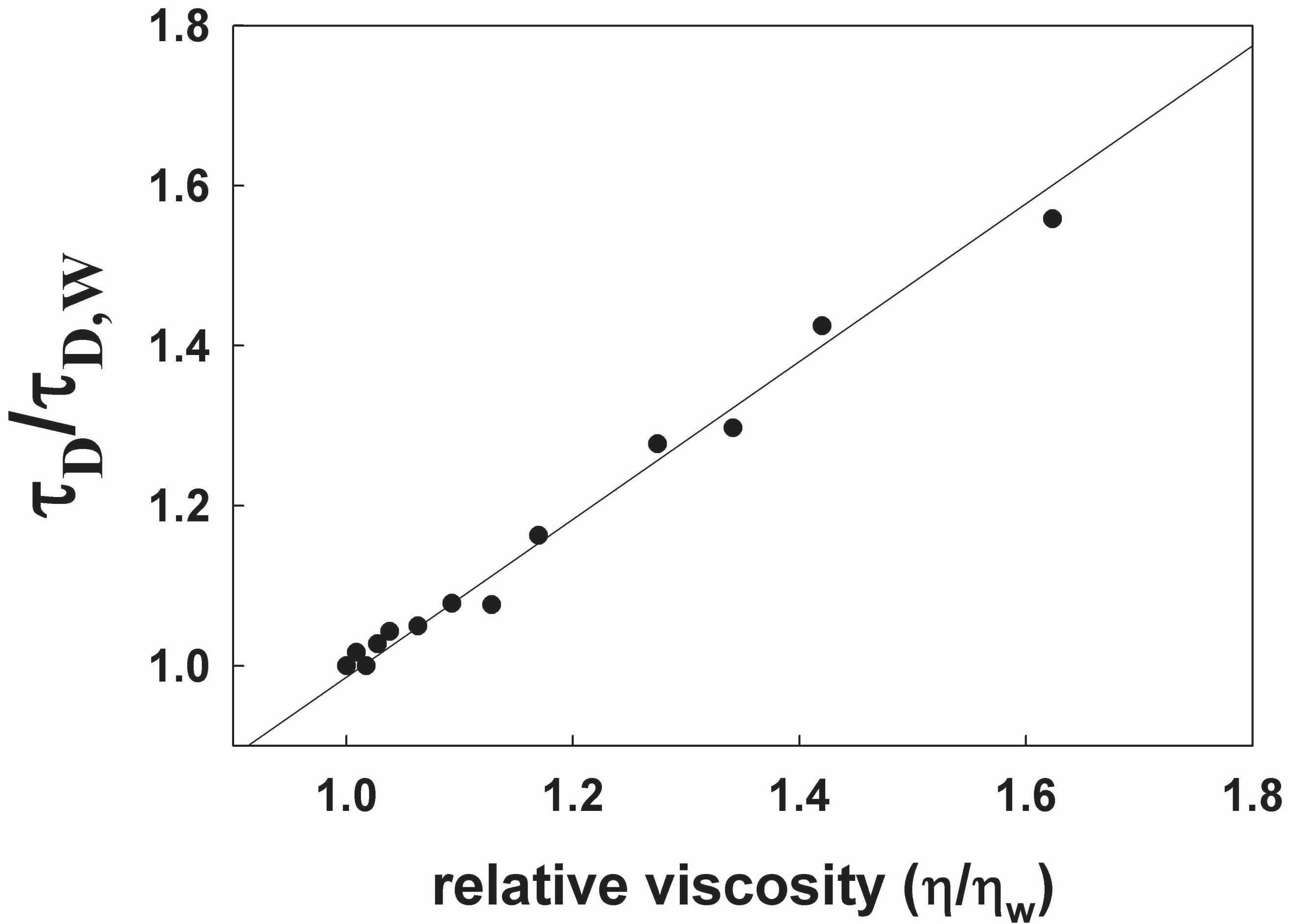

**Figure 3**

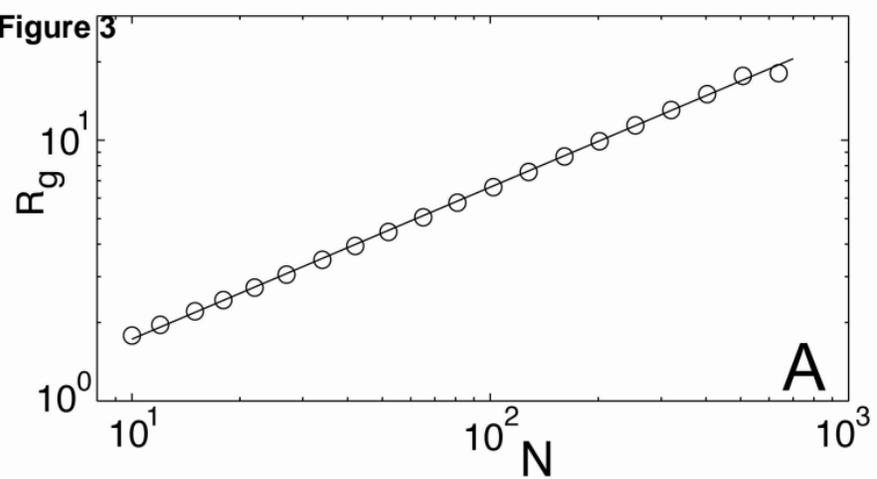
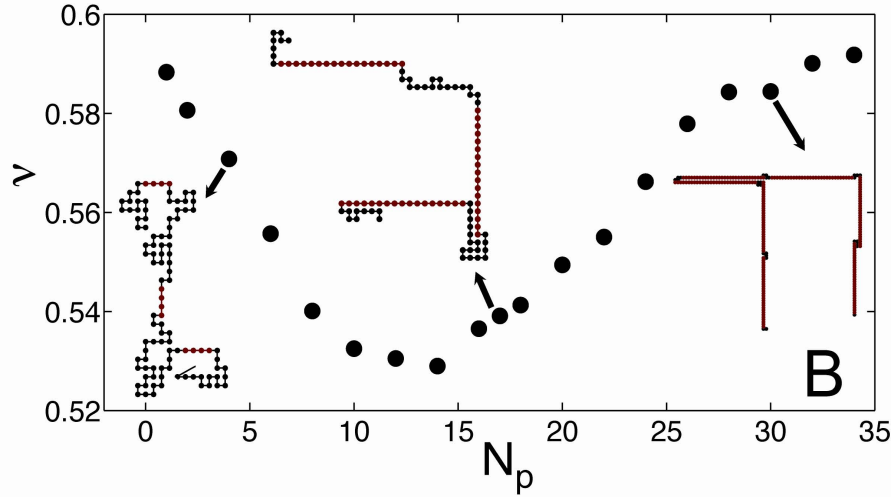
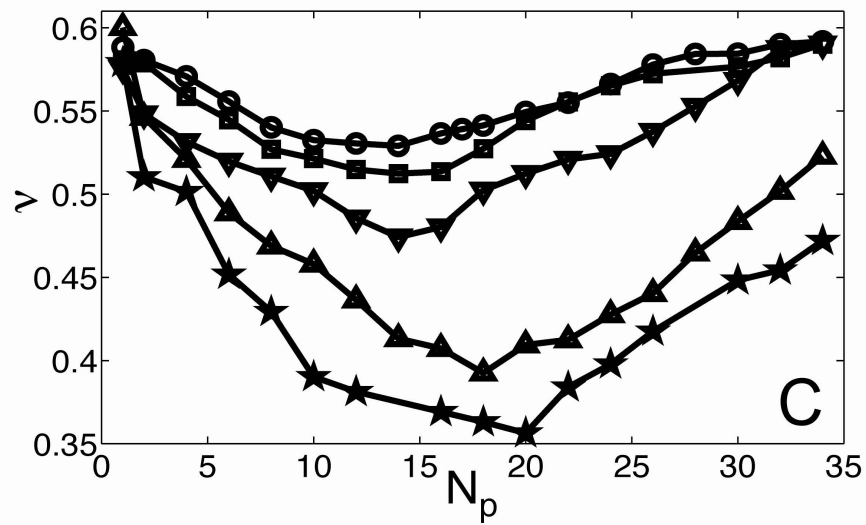
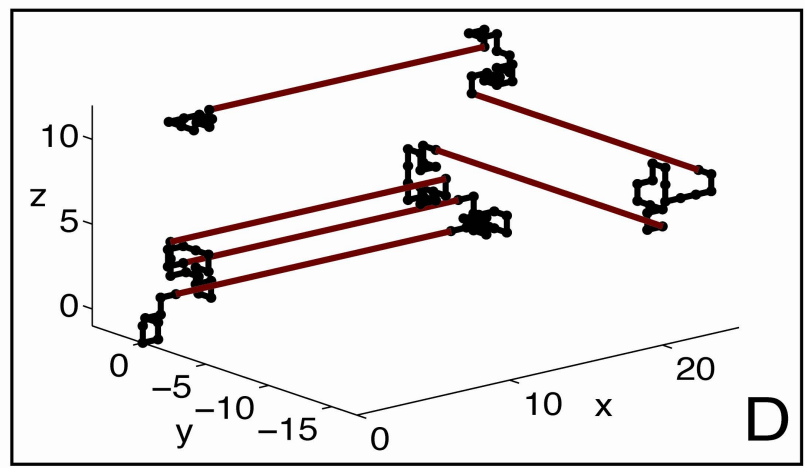